\documentclass[review]{elsarticle}

\usepackage{graphicx}
\usepackage{dcolumn}
\usepackage{pifont}
\usepackage{bm}
\usepackage{multirow}
\usepackage{amsmath}
\usepackage{float}
\usepackage{txfonts}
\usepackage[version=3]{mhchem}

\usepackage[dvips]{color}

\journal{Journal of Power Sources}

\begin{document}
\title{Revealing the formation and electrochemical properties of bis(trifluoromethanesulfonyl) imide intercalated graphite with first-principles calculations}

\author[kimuniv-m]{Chol-Jun Yu\corref{cor}}
\ead{ryongnam14@yahoo.com}
\author[kimuniv-m]{Un-Song Ri}
\author[kimuniv-m]{Gum-Chol Ri}
\author[kimuniv-m]{Jin-Song Kim}

\cortext[cor]{Corresponding author}

\address[kimuniv-m]{Computational Materials Design (CMD), Faculty of Materials Science, Kim Il Sung University, \\ Ryongnam-Dong, Taesong District, Pyongyang, Democratic People's Republic of Korea}

\begin{abstract}
Graphite has been reported to have anion as well as cation intercalation capacities as both cathode and anode host materials for the dual ion battery. In this work, we study the intercalation of bis(trifluoromethanesulfonyl) imide (TFSI) anion from ionic liquid electrolyte into graphite with first-principles calculations. We build models for TFSI-C$_n$ compounds with systematically increasing unit cell sizes of graphene sheet and investigate their stabilities by calculating the formation energy, resulting in the linear decrease and arriving at the limit of stability. With identified unit cell sizes for stable compound formation, we reveal that the interlayer distance and relative volume expansion ratio of TFSI-C$_n$ increase as increasing the concentration of TFSI intercalate during the charge process. The electrode voltage is determined to be ranged from 3.8 V to 3.0 V at the specific capacity ranging from 30 mAh g$^{-1}$ to 54 mAh g$^{-1}$ in agreement with experiment. Moreover, a very low activation barrier of under 50 meV for TFSI migration and good electronic conductivity give a proof of using these compounds as a promising cathode. Through the analysis of charge transfer, we clarify the mechanism of TFSI-C$_n$ formation, and reveal new prospects for developing graphite based cathode.
\end{abstract}


\maketitle

\section{Introduction}
Since the issue of global warming and climate change, originated from human activities of massively mining and consuming fossil fuels, is one of the greatest challenges of our time~\cite{Paris}, clean energy technologies are becoming increasingly important. More and more peoples are interested in the clean energy sources, such as wind and solar power, and electric vehicles or hybrid electric vehicles. In such intermittent energy systems and electrified drive transporters, electrochemical energy storage devices like rechargeable batteries are indispensable components. Lithium ion batteries (LIBs) have been dominating battery markets for portable electronic devices since their commercialization in 1991, due to their high specific capacity (low weight) and high energy density (low volume). However, the growing demand of LIBs and their introduction into vehicles have raised concerns about the low abundance and thus high cost of lithium resources~\cite{Tarascon10,Goodenough10}, together with other technical problems of safety, minimum charging time and cycle life~\cite{HZhang}. In recent years, therefore, numerous research studies have been made on alternatives with the same operational mechanism to LIBs based on high abundant and low cost elements such as sodium~\cite{JYHwang,Yabuuchi14,Sawicki} and potassium~\cite{Pramudita,BJi} towards a large-scale energy storage device with a reasonable cost.

Electrodes (cathode and anode) are the key components that determine the performance of alkali ion batteries, where alkali cations (\ce{Li+}, \ce{Na+}, \ce{K+}) intercalate/de-intercalate into/from the electrodes during the charge/discharge process. The standard anode widely used in commercial alkali ion batteries might be graphite because of its well-layered structure, composed of two-dimensional (2D) graphene layers bound through the weak van der Waals (vdW) attraction, which allows easy formation of graphite intercalation compounds (GICs)~\cite{Dresselhaus}. As electron donors, alkali metals (AMs) can form donor-acceptor type AM-GICs with the ionic bonding between AM and graphene layer, of which strength decreases as decreasing the atomic number (increasing the electropositivity) from Cs to Na, with the exception of Li due to the strong covalent bonding of Li$-$C~\cite{Moriwake,Nobuhara,ZWang}. From the theoretically calculated formation energies, it was found that for the stage 1 GICs the stability is in the order of \ce{KC6} $>$ \ce{LiC6} $>$ \ce{NaC6}, while \ce{NaC6} is thermodynamically unstable~\cite{Moriwake}. Moreover, the redox potential for K/\ce{K+} ($-2.93$ V) {\it vs.} standard hydrogen electrode is lower than for Na/\ce{Na+} ($-2.71$ V) and close to Li/\ce{Li+} ($-3.04$ V), suggesting higher cell voltage for K-ion battery (KIB) compared to Na-ion battery (NIB) and similar value to LIB. On the other hand, the larger radius ion was found to diffuse more smoothly in graphite due to the activation barriers in the order of Li $>$ Na $>$ K~\cite{Nobuhara,yucj06}. Together with a relatively high abundance, a low cost and environmental friendliness of potassium, these indicate more potentiality of KIB rather than NIB and LIB.

For KIB to become competitive with LIB in battery markets, however, suitable cathode material with a high capacity and a high cell voltage should be developed. As for LIBs and NIBs, many kinds of materials, such as layered metal oxides~\cite{Vaalma,XRen} and polyanionic compounds~\cite{JHan,Recham}, have shown promising properties. Alternatively, based on the unique amphoteric character of graphite that can accommodate both cations (electron donor) and anions (acceptor), graphite was suggested to be used as the cathode host as well, namely dual-ion battery (DIB) or dual-graphite battery (DGB), where AM cations and complex anions simultaneously intercalate/de-intercalate into the graphite based anode and cathode during the charge/discharge process~\cite{YWang,Lebdeh,Carlin,Seel,Schmuelling,Beltrop2,Rothermel,Meister}. Here, the anions are from ionic liquids used as electrolyte in ion batteries, which are typically hexafluorophosphate (\ce{PF6-})~\cite{Seel}, perchlorate (\ce{ClO4-})~\cite{Santhanam}, fluorosulfonyl imide (FSI, \ce{[N(SO2F)2]-}), and bis(trifluoromethanesulfonyl) imide (TFSI, \ce{[N(SO2CF3)2]-})~\cite{Fujii,Umebayashi,Henderson,Henderson2,Bhatt,Siqueira,Borodin1,Borodin2,Nicotera}. Recently, Beltrop {\it et al.}~\cite{Beltrop1} reported a novel potassium-based DGB (K-DGB), composed of $N$-butyl-$N$-methyl-pyrrolidinium TFSI (\ce{Pyr14+TFSI-}) + 0.3 M potassium TFSI (\ce{K+TFSI-}) + 2 wt\% ethylene sulfite (ES) as electrolyte and graphite as both the anode and cathode hosts, showing a relatively high reversible capacity of $\sim$230 mAh g$^{-1}$ and a high potential range from 3.4 V to 5.0 V {\it vs.} K/\ce{K+}. This requires to reveal the mechanism of formation and electrochemical properties of TFSI-GICs by applying the theoretical and computational method. To the best of our knowledge, however, there is no theoretical study on TFSI-GICs, although some simulation works for TFSI molecule~\cite{Fujii,Umebayashi,Siqueira,Borodin1} and first-principles work for other anion GICs such as \ce{PF6}- and \ce{ClO4}-GICs~\cite{Tasaki} have been reported.

In this study, we apply the first-principles method within the density functional theory (DFT) framework to TFSI-GICs, together with graphite and TFSI molecule, aiming to reveal energetics, structures, electrochemical properties, and formation mechanism. Supercells of TFSI-C$_n$, where the carbon atom number $n$ is determined from unit cells of graphene sheet with various sizes, are built, and their formation and interlayer binding energies are calculated. The electrode voltage, activation barrier for TFSI migration, and electronic conductance are considered. To obtain insights for GIC formation, we perform an analysis of electronic density difference and atomic charge.

\section{Computational methods}
The DFT calculations were performed with the pseudopotential plane wave method as implemented in Quantum ESPRESSO code (version 5.3.0)~\cite{QE}, using the ultrasoft pseudopotentials provided in the package.~\footnote{We used the Vanderbilt-type ultrasoft pseudopotentials C.pbe-van\_ak.UPF, N.pbe-van\_ak.UPF, S.pbe-van\_bm.UPF, O.pbe-van\_ak.UPF, and F.pbe-n-van.UPF, which are provided in the package.} Here, valence electrons of C 2s$^2$2p$^2$, N 2s$^2$2p$^3$, O 2s$^2$2p$^4$, S 3s$^2$3p$^4$, and F 2s$^2$2p$^5$ were explicitly considered. For exchange-correlation (XC) interaction between the valence electrons, Perdew-Burke-Ernzerhof (PBE)\cite{PBE} functional within the generalized gradient approximation (GGA) was used, and using the vdW-DF2 functional~\cite{vdwDF2}, the dispersive energy was added. The cutoff energies were set to be 40 Ry for a plane wave basis set and 400 Ry for an electronic density. An isolated TFSI molecule was simulated using the cubic supercell with a lattice constant of 17 \AA, which is large enough to prevent the artificial interaction between the neighbouring images. Only $\Gamma$ point in the Brillouin zone (BZ) was used for this isolated molecule. To make modelling stage 1 GICs, TFSI-C$_n$, one TFSI molecule is placed between the graphene sheets in $AA$-stacked graphite with various cell sizes of $(3\times 3)$, $(4\times 3)$, $(4\times 4)$, $(5\times 4)$, $(5\times 5)$, $(6\times 5)$ and $(6\times 6)$, which give the carbon atom number as $n$ = 18, 24, 32, 40, 50, 60 and 72, respectively. The special $k$-points for the BZ sampling were constructed using the $(3\times 3\times 3)$ and $(3\times 3\times 5)$ Monkhorst-Pack meshes for the $(3\times 3)$ $-$ $(5\times 4)$ and $(5\times 5)$ $-$ $(6\times 6)$ cells for atomic relaxations, while denser $k$-points using the $(8\times 8\times 8)$ and $(8\times 8\times 10)$ meshes for DOS calculations. These computational parameters of cutoff energy and $k$-point mesh could produce the absolute total energy convergence better than 1 meV per atom. In the atomic relaxations, the forces on each atom converged to within $5\times10^{-4}$ Ry Bohr$^{-1}$.

In order to study of the TFSI intercalation into graphite, we assumed the following process~\cite{Tasaki},
\begin{gather}
\ce{TFSI-}-e \rightarrow \ce{TFSI} \\
\ce{TFSI}+\ce{C}_n \rightarrow \text{TFSI-C}_n \label{eq2}
\end{gather}
The formation energy of TFSI-C$_n$ compound per formula unit (four carbon atoms) can be calculated as follows,
\begin{equation}
E_f = \frac{4}{n}\left(E({\text{TFSI-C}_n}) - \frac{n}{4}E(\text{graphite}) - E(\text{TFSI})\right) \label{eq_Ef}
\end{equation}
where $E({\text{TFSI-C}_n})$, $E(\text{graphite})$, and $E(\text{TFSI})$ are the DFT total energies of TFSI-C$_n$ supercell, graphite unit cell, and isolated TFSI supercell, respectively. The negative formation energy indicates that the GIC is thermodynamically stable, {\it i.e.} exothermic formation of the GIC from graphite and TFSI molecule. The interlayer binding energy (or exfoliation energy) per carbon atom can be calculate as follows~\cite{yucj14},
\begin{equation}
E_b = \frac{1}{n}\left(E({\text{TFSI-C}_n})_{(d_i=d_e)} - E({\text{TFSI-C}_n})_{(d_i=d_\infty)}\right)\label{eq_Eb}
\end{equation}
where $d_i$ is the interlayer distance, $d_e$ the equilibrium interlayer distance, and $d_\infty$ might be 20 \AA~over which the total energy is little changed. To calculate the activation barrier for TFSI migration inside graphite, the climbing image nudged elastic band (CI-NEB) method~\cite{NEB} was applied to TFSI-\ce{C50} compound using seven image points and convergence threshold of 0.05 eV \AA$^{-1}$ for the force on band orthogonal to path. The supercell dimensions were fixed at the optimized one, while all atoms were relaxed.

\section{Results and discussion}
\subsection{Structure and energetics}
In compliance with the well-known fact that standard XC functionals within GGA or LDA poorly describe graphite and other carbon layered materials, and therefore, vdW functional should be adopted~\cite{yucj06,yucj14,yucj04,Tsai}, we first assessed the reliability of various vdW functionals by calculating lattice constants of graphite. Table S1$\dag$ shows that all PBE + vdW functionals adopted in this work, even PBE itself, reproduced the experimental in-plane lattice constant (2.461 \AA~\cite{boettiger}), governed by the \ce{C-C} covalent ($\sigma$ and $\pi$) bonding, while vdW-DF2 functional~\cite{vdwDF2} (relative error 1.64\%) yielded the best agreement with the experimental interlayer distance (3.353 \AA~\cite{boettiger}). As the interlayer distance is perpendicular to the graphene sheet and along the direction of the vdW interaction, vdW-DF2 functional was decided to be the most reliable one for graphite and further GICs. For a check of the stability of graphite, the interlayer binding energy was calculated to be $-$78 meV per atom. When compared with the experimental values in the range of $-$35 $\sim$ $-$52 meV~\cite{graphitexf1,graphitexf2}, our calculation gave a slight overestimation, but close to other theoretical results, such as $-$69 meV with the inclusion of vdW energy by Tasaki~\cite{Tasaki}, and $-$61 $\sim$ $-$74 meV with the combination of DFT and perturbation theory by Dapper {\it et al.}~\cite{graphitexf3}.
\begin{table*}[!th]
\centering
\caption{\label{tab_tfsi}Bond length, bond angle, dihedral angle and molecular volume of TFSI in isolated state and TFSI-C$_n$ compounds ($n$ = 18, 24, 32, 40, 50), determined using the PBE + vdW-DF2 functional. Molecular volume is calculated based on the Connolly surface.}
\begin{tabular}{ccccccc}
\hline
          &       & \multicolumn{5}{c}{TFSI-C$_n$} \\
 \cline{3-7}
          & TFSI  & 18    & 24    & 32    & 40    & 50 \\
\hline
\multicolumn{7}{l}{Bond length (\AA)} \\
S$-$N     &1.642  &1.644  &1.623  &1.618  &1.616  &1.614 \\
S$-$O     &1.463  &1.453  &1.451  &1.449  &1.448  &1.447 \\
S$-$C     &2.027  &1.901  &1.905  &1.901  &1.901  &1.900 \\
C$-$F     &1.343  &1.370  &1.377  &1.379  &1.381  &1.383 \\
\multicolumn{7}{l}{Bond angle (degree)} \\
S$-$N$-$S &130.06 &126.07 &126.19 &125.64 &125.68 &125.26 \\
N$-$S$-$O &112.28 &116.52 &115.57 &115.73 &115.88 &115.98 \\
          &105.61 &106.55 &107.67 &108.09 &108.21 &108.51 \\
N$-$S$-$C &107.81 &100.60 &103.46 &103.33 &103.31 &103.20 \\
O$-$S$-$C &105.44 &105.73 &104.23 &104.02 &103.86 &103.69 \\
O$-$S$-$O &119.59 &119.58 &119.65 &119.53 &119.51 &119.46 \\
S$-$C$-$F &105.97 &109.38 &110.45 &110.90 &111.12 &111.33 \\
F$-$C$-$F &112.74 &109.56 &108.47 &108.00 &107.77 &107.54 \\
\multicolumn{7}{l}{Dihedral angle (degree)} \\
S$-$N$-$S$-$C&55.74 &84.94 &88.28 &87.77  &87.87  &87.70 \\
\multicolumn{7}{l}{Molecular volume (\AA$^3$)} \\
          &164.4  &161.6  &163.7  &162.6  &163.7  &163.3 \\
\hline
\end{tabular}
\end{table*}

We then considered the isolated TFSI molecule, as assumed in this work according to Eq.~\ref{eq2}, although TFSI exists in the anion state in the form of room-temperature ionic liquids, such as \ce{Li+TFSI-}~\cite{Umebayashi,Henderson,Borodin2}, \ce{K+TFSI-}~\cite{Beltrop1} and \ce{Pyr14+TFSI-}~\cite{Henderson,Siqueira,Borodin1,Borodin2,Nicotera,Beltrop1}, used as electrolyte for polymer batteries and alkali ion batteries. The molecular structure was optimized after conformation search and analysed with a comparion to those of TFSI-C$_n$ compounds. In addition, the electrostatic potential and frontier molecular orbitals such as the highest occupied molecular orbitals (HOMO) and the lowest unoccupied molecular orbitals (LUMO) were calculated. Since TFSI has the molecular structure of \ce{F3C-S(O2)-N-S(O2)-CF3} and the terminal \ce{CF3} group can rotate along the \ce{S-N} bond to give rotational isomers, there are typically two conformations according to the dihedral \ce{C-S-N-C} angles; $C_1$ ($cis$) and $C_2$ ($trans$)~\cite{Fujii,Umebayashi}. In accordance with the previous theoretical works employing the Gaussian orbital method~\cite{Fujii,Umebayashi} and force-field molecular dynamics~\cite{Siqueira,Borodin1}, it turned out from conformation search with the systematic grid scan method that the $C_1$ conformer is the energetically lowest conformer.

\begin{figure}[!b]
\centering
\includegraphics[clip=true,scale=0.16]{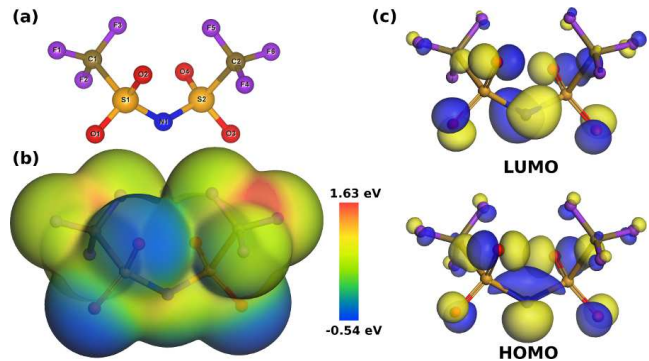}
\caption{\label{fig_tfsi}(a) The molecular structure of TFSI molecule with the lowest energy conformation, optimized using the PBE + vdW-DF2 functional, (b) 3D volumetric map of its electrostatic potential, and (c) isosurface plot of frontier molecular orbitals including HOMO and LUMO.}
\end{figure}
This conformer was placed within the cubic supercell with a lattice constant of 17~\AA~to perform atomic relaxation using the vdW-DF2 functional. Effective screening medium method~\cite{esm} was applied to this isolated molecule as implemented in the code. The optimized molecular structure of TFSI molecule is depicted in Figure~\ref{fig_tfsi}(a), and the chemical bonding properties such as bond length, bond angle and dihedral angle are presented in Table 1. When compared with those by Gaussian method~\cite{Fujii,Umebayashi}, the bond lengths were calculated to be slightly larger in overall, especially \ce{S-C} bond length, possibly due to the inclusion of vdW interaction. To obtain qualitative insight for the electrochemical reduction stability, we present electrostatic potential maps in Figure~\ref{fig_tfsi}(b). The regions of lowest electrostatic potential (Figure~\ref{fig_tfsi}(b), bluest region) are found in the vicinity of O atoms, whilst the regions of highest electrostatic potential (Figure~\ref{fig_tfsi}(b), reddest region) around the F, C and S atoms, indicating the electrochemical reduction will be occurred by oxygen atoms. We also plotted the frontier molecular orbitals including HOMO and LUMO, as presented in Figure~\ref{fig_tfsi}(c). It is interesting that around S atoms neither HOMO nor LUMO is found, indicating a different role of S atom from other atoms. In addition, the intra-molecular charge separation upon excitation can not be said to occur due to no clear separation between HOMO and LUMO distribution regions.

\begin{figure}[!t]
\centering
\includegraphics[clip=true,scale=0.49]{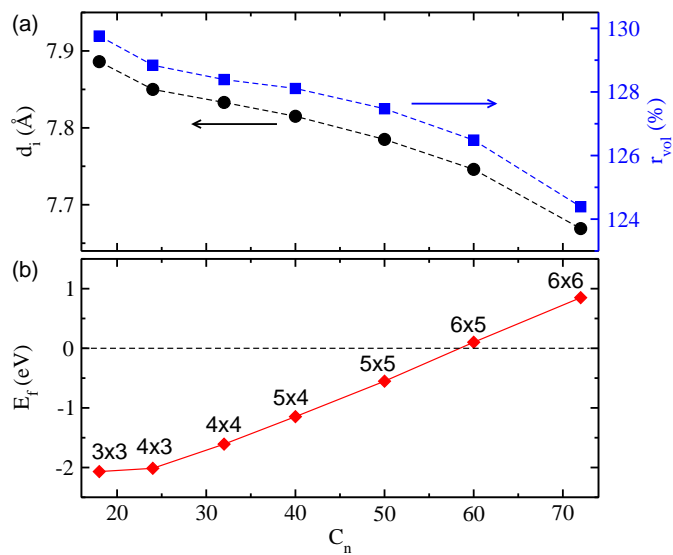}
\caption{\label{fig_lengef}(a) The interlayer distance ($d_i$) and relative volume expansion ratio ($r_\text{vol}$), and (b) formation energy per formula unit (4 carbon atoms) ($E_f$), as functions of carbon atom number (C$_n$) of graphene sheet in TFSI-C$_n$ compounds. Cell sizes of graphene sheets are marked in (b).}
\end{figure}
With the structural and energetic characteristics of graphite and TFSI molecule, next, we applied the same DFT scheme to the TFSI-C$_n$ compounds under study. Due to a certain amount of TFSI molecular volume, the minimum cell size of 2D graphene sheet should be determined to intercalate it into graphite. The molecular volume of TFSI was calculated to be 152.9 or 164.4 \AA$^3$ based on the vdW or Connolly surface, which gives a radius of 5.35 or 5.48 \AA~when reduced to the sphere. Then, the $(3\times3)$ graphene sheet with a cell width of 7.41 \AA~should be minimal for TFSI to be fully placed inside the sheet. On the other hand, it was found that the stability of ternary GICs with the co-intercalate consisted of AM atom and diglyme solvent molecule, as estimated by the formation energy, reduces as decreasing the concentration of intercalate, and even the GIC becomes unstable under a certain value of concentration~\cite{yucj14}. Thus, we calculated the formation energies of TFSI-C$_n$ as systematically increasing the number of carbon atom in graphene sheet, {\it i.e.}, increasing the cell size of graphene sheet from $(3\times3)$ ($n$ = 18) to $(6\times6)$ ($n$ = 72) (decreasing the TFSI concentration). Figure~\ref{fig_lengef} shows the formation energy as a function of carbon atom number, together with the interlayer distances and volume expansion ratios relative to pristine graphite. It is seen that the formation energy has the minimum value as $-$2.07 eV in the \ce{C18} compound, and increases linearly as increasing the carbon atom number from $n$ = 24 (similar value of $-$2.01 eV to \ce{C18}), becoming positive over $(6\times5)$ cell ($n$ = 60). This indicates that TFSI-C$_n$ formation over $n$ = 60, from graphite and TFSI molecule, is not suitable thermodynamically, and therefore, TFSI-C$_n$ with $n$ = 18, 24, 32, 40, and 50 were considered for further calculation and analysis.

\begin{figure}[!t]
\centering
\includegraphics[clip=true,scale=0.27]{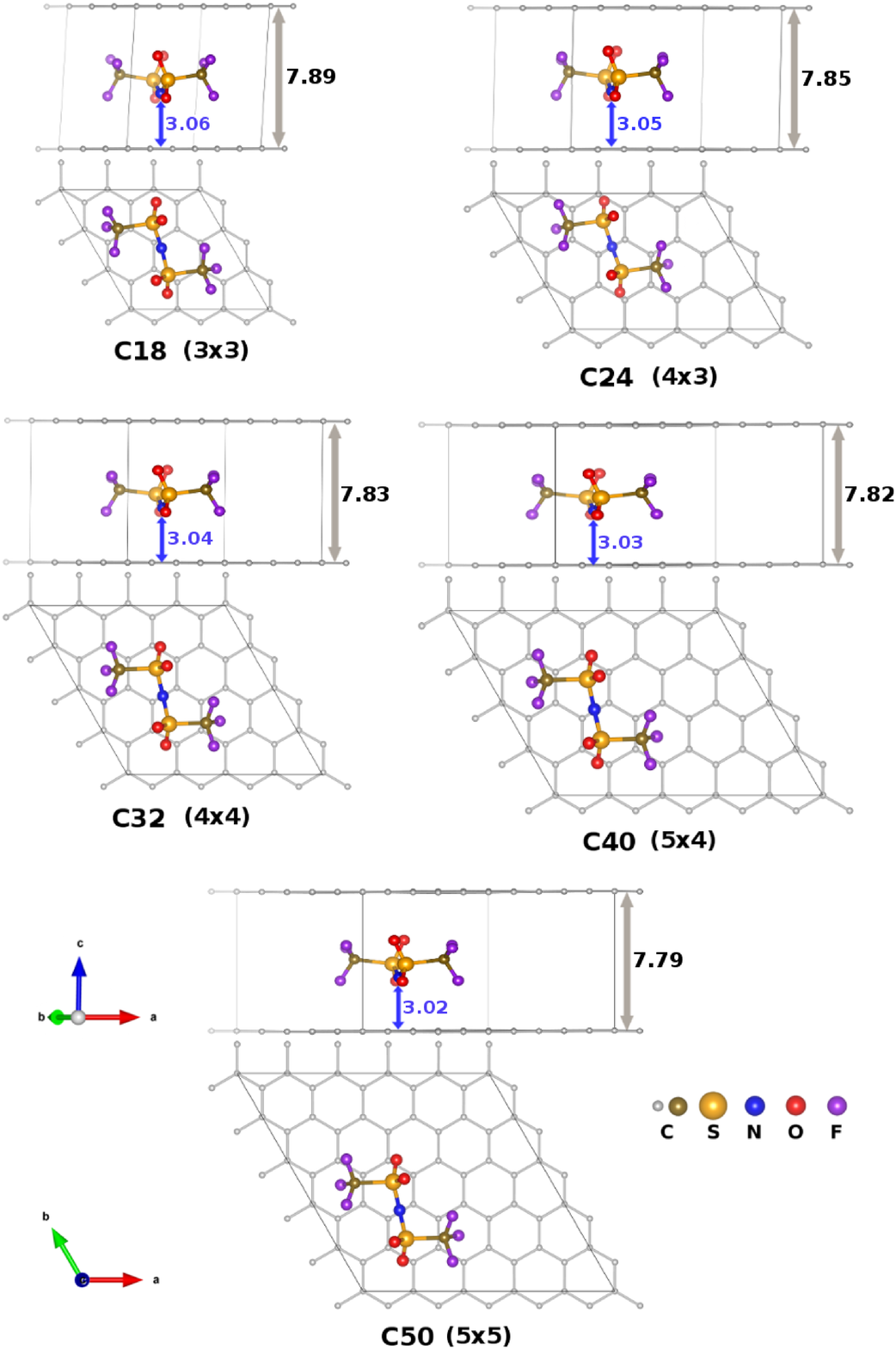}
\caption{\label{fig_struct}Side and top views of supercells of TFSI-C$_n$ compounds ($n$ =18, 24, 32, 40, 50), optimized using the PBE + vdW-DF2 functional. Interlayer distance and height of N atom to graphene sheet in the unit of \AA~are marked.}
\end{figure}
Figure~\ref{fig_struct} shows the atomistic structures of these GICs, optimized with PBE + vdW-DF2 functional. It was shown that the interlayer distance has the maximum value in TFSI-\ce{C18} as 7.89 \AA, determined through the variable cell optimization, and decreases as a linear function of carbon atom number, while the relative volume expansion ratio also decreases from 130\% (\ce{C18}) to 127\% (\ce{C50}) (see Figure~\ref{fig_lengef}(a)). The height of N atom to the bottom graphene sheet decreases monotonically as increasing the carbon atom number. When compared with the experimental values of $d_i\approx8.21$ \AA~and $r_\text{vol}\approx140$\%~\cite{Schmuelling,Beltrop1}, our calculation gave underestimation in accordance with the above-mentioned slight overestimation of interlayer binding. In Table~\ref{tab_tfsi}, we present the bond length, bond angle, dihedral angle and molecular volume of TFSI molecule intercalated into graphite, in comparison with those in isolated state. It was observed that the bond lengths of \ce{S-N}, \ce{S-O} and \ce{S-C} lessens upon the intercalation of TFSI into graphite, whilst the \ce{C-F} bond lengthens. The bond angles changes also systematically in overall, and in particular, the dihedral angle of \ce{S-N-S-C} changed greatly from 56 degree in isolated state to over 84 degree in TFSI-C$_n$ GICs. These correlate well with the molecular volume contraction from 164.4 \AA$^3$ in isolated state to 161.6 $\sim$ 163.7 \AA$^3$ in GICs. Note that the contraction of TFSI molecular volume upon intercalation is the most remarkable in the $(3\times3)$ cell (\ce{C18}), the highest concentration of intercalate.

\begin{figure}[!t]
\centering
\includegraphics[clip=true,scale=0.52]{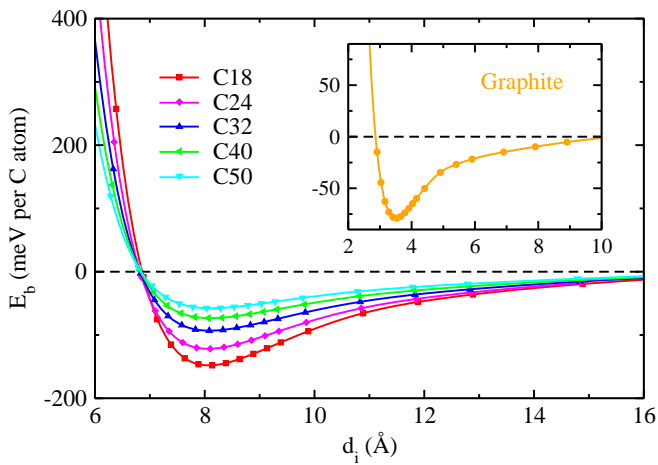}
\caption{\label{fig_exfol}The interlayer bind energy energy per carbon atom in TFSI-C$_n$ compounds with $n$ = 18, 24, 32, 40, and 50, calculated using the PBE + vdW-DF2 functional. Inset shows one of graphite.}
\end{figure}
As the another energetic property of GICs in addition to formation energy, we calculated the interlayer binding energy (Eq.~\ref{eq_Eb}), representing the strength of binding between graphene sheets, as shown in Figure~\ref{fig_exfol}. The interlayer binding energy of TFSI-C$_n$ compounds decreases in magnitude as increasing the cell size of graphene sheet, indicating a decrease of binding strength, which correlates well with the tendency of formation energy. Moreover, those with $n$ = 18 ($-$148 meV), 24 ($-$122 meV), 32 ($-$93 meV) are bigger than that of graphite ($-$78 meV), indicating that the GIC formation in these cases enhances the interlayer binding compared with graphite due to a change of bonding from vdW characteristics to ionic one. If the number of carbon atom is over 32, {\it i.e.} $n$ =40 ($-$73 meV) and 50 ($-$58 meV), however, those are smaller than that of graphite, indicating a weakening of interlayer binding.

\subsection{Electrochemical characteristics}
In this subsection, we consider typical electrochemical characteristics of TFSI-C$_n$ compounds, such as electrode voltage, intercalate diffusion and electronic conductivity. As the electrode voltage is one of the most crucial electrochemical properties, better cathode material should possess higher electrode voltage. Providing that, during the charge process in DIB, TFSI-GIC transforms from the compound of higher carbon atom number (lower concentration of TFSI intercalate) to that of lower one (higher concentration of TFSI intercalate), we can change the notation of TFSI-GICs from TFSI-C$_n$ to TFSI$_{x_i}$-C$_{60}$, where $x_i=60/n$ using C$_{60}$ GIC as the starting compound. Then, using the total energies of the compounds, the electrode voltage can be calculated as follows,
\begin{equation}
 V_\text{el}=-\frac{x_iE(\ce{TFSI}_{x_i}\text{-C}_{60})-x_jE(\ce{TFSI}_{x_j}\text{-C}_{60})-(x_i-x_j)E(\ce{TFSI})}{(x_i-x_j)e}
\end{equation}
where $e$ is the elementary charge of electron. It should be noted that, although TFSI-\ce{C60} compound was estimated to be unstable, it can be used as a reference compound due to its low positive formation energy as 0.10 eV (Figure~\ref{fig_lengef}(b)).

\begin{figure}[!t]
\centering
\includegraphics[clip=true,scale=0.14]{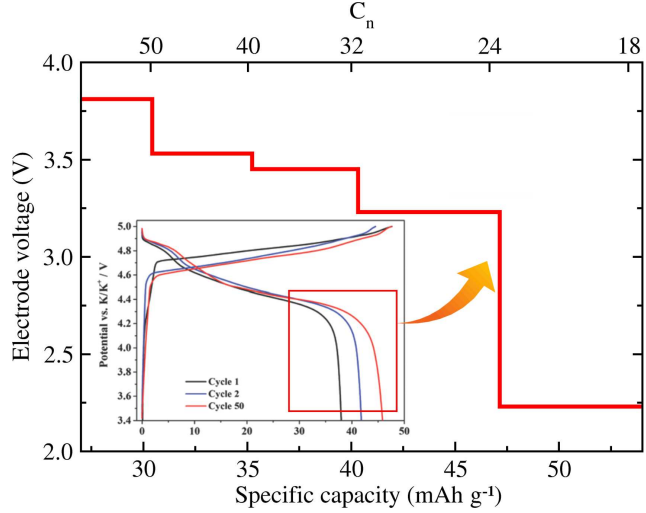}
\caption{\label{fig_vel}The electrode voltage step as increasing the carbon atom number and accordingly specific capacity. Inset shows experimental electrode potential obtained by Beltrop {\it et al.}~\cite{Beltrop1}, with the red rectangle indicating the range corresponding to calculation.}
\end{figure}
\begin{table}[!b]
\caption{\label{tab_sum}Overview of interlayer distance ($d_i$), relative volume expansion ratio ($r_\text{vol}$), formation energy per formula unit ($E_f$), interlayer binding energy per carbon atom ($E_b$), specific capacity (SC) and average electrode voltage ($V_\text{el}$) in TFSI-C$_n$ GICs with $n$ from 18 to 50.}
\begin{tabular}{ccccccc}
\hline
     & $d_i$ & $r_\text{vol}$ & $E_f$ & $E_b$ & SC   & $V_\text{el}$ \\
C$_n$& (\AA)        & (\%)           & (eV)  & (meV) & (mAh g$^{-1}$) & (V) \\
\hline
18 & 7.89 & 129.76 & $-$2.07 & $-$148 & 54.0 & 3.00 \\
24 & 7.85 & 128.44 & $-$2.01 & $-$122 & 47.2 & 3.42 \\
32 & 7.83 & 128.39 & $-$1.61 & $-$93  & 40.3 & 3.56 \\
40 & 7.82 & 128.11 & $-$1.15 & $-$73  & 35.2 & 3.64 \\
50 & 7.79 & 127.47 & $-$0.55 & $-$58  & 30.4 & 3.81 \\
\hline
\end{tabular}
\end{table}
Figure~\ref{fig_vel} shows the electrode voltage step as progressing the charge process, {\it i.e.} increasing the number of carbon atom of graphene sheet and accordingly the specific capacity. It was found that as decreasing the carbon atom number, {\it i.e.} cell size of graphene sheet, the specific capacity increases from 30.4 mAh g$^{-1}$ (TFSI-\ce{C50}) to 54.0 mAh g$^{-1}$ (TFSI-\ce{C18}), yielding the voltage steps from 3.81 V to 2.23 V. When compared with the experimental voltage range from 5.0 V to 3.4 V {\it vs.} K/\ce{K+} and specially from 4.4 V to 3.4 V corresponding to the capacity range from 30  to 45 mAh g$^{-1}$, our calculated values were in reasonable agreement with the experiment. Table~\ref{tab_sum} presents an overview of structural, energetic, and electrochemical characteristics of TFSI-C$_n$ compounds with $n$ from 18 to 50, including interlayer distance, relative volume expansion ratio, formation energy, interlayer binding energy, specific capacity and average electrode voltage.

\begin{figure}[!t]
\centering
\includegraphics[clip=true,scale=0.11]{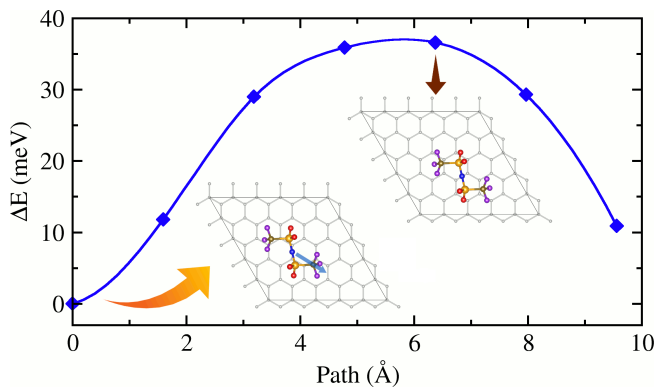}
\caption{\label{fig_neb}The activation barrier for TFSI migration inside the space between graphene sheets along the path across one carbon hexagon. Top views of starting and transition states are presented.}
\end{figure}
The mobility of intercalate inside space formed between graphene sheets can determine the rate capability and cycling stability of graphite based electrode. In this respect, we calculated the activation barriers for TFSI migrations along possible paths across one carbon hexagon by applying the CI-NEB method~\cite{NEB} to TFSI-\ce{C50} GICs. The activation barrier was calculated to be very low like at most $\sim$35 meV, as shown in Figure~\ref{fig_neb} (see Figure S1-S3$\dag$ for those along other paths). When compared with those for alkali cation (0.2 - 0.4 eV)~\cite{Nobuhara} and co-intercalate composed of AM and solvent molecule (0.1 $\sim$ 0.6 eV)~\cite{yucj06,yucj14} migrations inside graphite, this is surprisingly so low that TFSI can migrate almost freely with a very long diffusion length. This suggests a very fast charging time and very long cycling life in DIB. However, it should be noted that much higher activation energy can be required for the first anion uptake, due to the initial widening of the interlayer distance between graphene sheet against the attracting vdW forces.

\begin{figure}[!t]
\centering
\includegraphics[clip=true,scale=0.55]{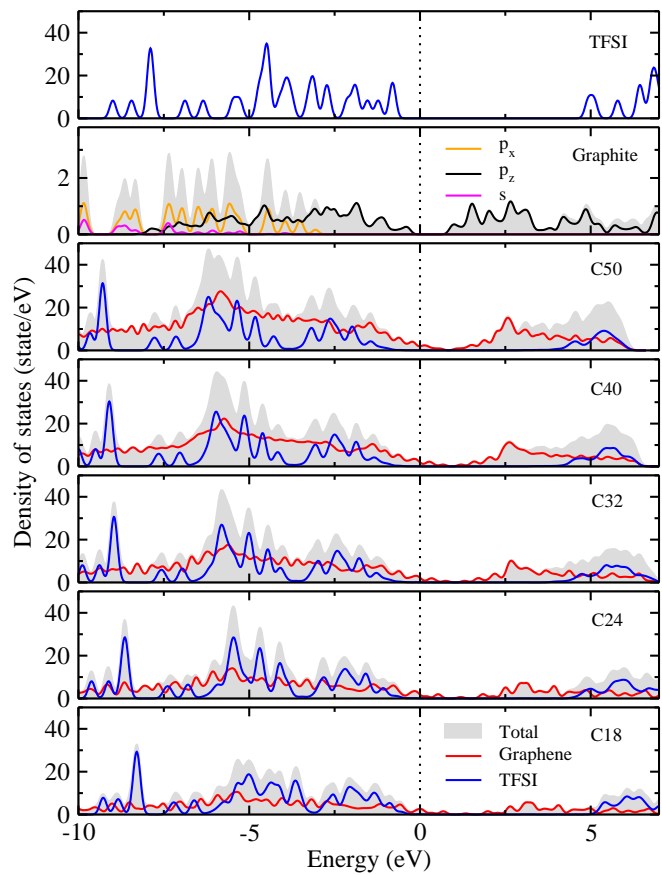}
\caption{\label{fig_dos}The electronic density of states (DOS) in isolated TFSI molecule, graphite, and TFSI-C$_n$ compounds with $n$ = 18, 24, 32, 40, and 50, calculated using the PBE + vdW-DF2 functional. In addition to total DOS, angular momentum dependent projected DOS for the case of graphite and partial DOS of TFSI and graphene sheet for the cases of GICs are presented. Fermi energy is set to zero as indicated by dotted vertical line.}
\end{figure}
The electronic conductivity is also an important property for understanding the battery operation. As we assumed for \ce{TFSI-} anion intercalation into graphite, the \ce{TFSI-} anion releases an electron before stating the intercalation, and the electron can be transmitted through GICs to the current collector. Therefore, TFSI-C$_n$ compound should be an electronic conductor like many other electrode materials. The electron conductivity of materials can be judged qualitatively by analysing the electronic density of states (DOS). Figure~\ref{fig_dos} presents the total DOS in isolated TFSI molecule, graphite, and TFSI-C$_n$ compounds, with angular momentum projected DOS for the case of graphite and atom resolved partial DOS for the cases of GICs. As already being well-known, graphite was confirmed to have good in-plane electronic conductivity due to an existence of $p_z$ electron states over the Fermi level, but no inter-plane conductivity due to $s$ and $p_x$ or $p_y$ states away from the Fermi level. Upon TFSI intercalation into graphite, the in-plane conductivity of graphene sheet was observed to preserve well due to electronic states of graphene sheets around the Fermi level in TFSI-C$_n$ compounds. When increasing the concentration of TFSI intercalate (decreasing the carbon atom number), the unoccupied states of graphene are getting further away from Fermi level, indicating a reduction of electronic conductivity. In addition, the overlap of electronic states between graphene sheet and TFSI molecule was observed below and above the Fermi level, indicating hybridization between $\sigma$ bonding orbitals of graphene sheet and molecular orbitals of TFSI molecule.

\subsection{Electronic charge transfer}
\begin{figure*}[!t]
\centering
\includegraphics[clip=true,scale=0.35]{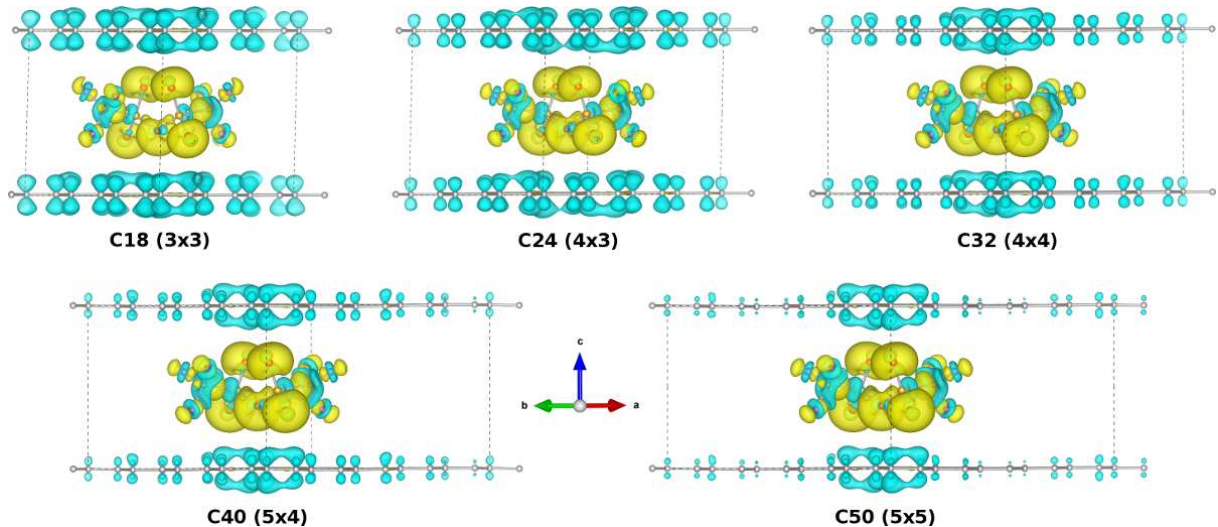}
\caption{\label{fig_chg}Isosurface plot of the electronic density differences at the value of 0.002 $|e|$ \AA$^{-3}$ upon TFSI intercalation into graphite forming TFSI-C$_n$ compounds with $n$ = 18, 24, 32, 40, and 50. Yellow colour is for positive value (electron gain), while cyan colour is for negative value (electron loss).}
\end{figure*}

Finally, we performed an analysis of electronic density difference and atomic charge to obtain insights into the charge transfer occurred upon TFSI intercalation into graphite forming TFSI-C$_n$ compounds and the chemical bonding between TFSI and graphene layer. This can be meaningful for understanding the mechanism of TFSI-GIC formation. The electronic density difference ($\Delta\rho$) is obtained as the difference between the electronic densities of the TFSI-C$_n$ compounds ($\rho(\ce{TFSI}\text{-C}_n)$) and those of the graphene sheet ($\rho(\text{C}_n)$) and TFSI molecule ($\rho(\ce{TFSI})$) as follows,
\begin{equation}
 \Delta\rho=\rho(\ce{TFSI}\text{-C}_n)-\rho(\text{C}_n)-\rho(\ce{TFSI})
\end{equation}
where the atomic positions are fixed as the optimized one.

Figure~\ref{fig_chg} shows the electronic density differences with an isosurface plot at the value of 0.002 $|e|$ \AA$^{-3}$. On condition that the positive value (yellow colour) indicates the electronic density accumulation and the negative value (cyan colour) the density depletion, it was found that the valence electronic density of the graphene layer has transferred almost totally to the TFSI molecule, and the extent of charge transfer decreases gradually as increasing the cell size of graphene layer due to an intuitive gradual reduction in the amount of charge depletion on the graphene layer. Nevertheless, the amount of electronic density difference around the carbon atoms facing the oxygen and nitrogen atoms of TFSI is kept almost to be constant. Moreover, the distribution of electronic density difference around the TFSI molecule also seems to be little changed. Inside the TFSI molecule, the major portion of charge accumulation is found around the O and N atoms, confirming the ionic character of the bonding between the nitrogen as well as oxygen atoms of TFSI molecule and the facing carbon atoms of graphene layer.

To quantitatively estimate the amount of transferred charge, the atomic charges calculated by using the L\"{o}wdin method were analysed. As shown in Table S2$\dag$, upon the intercalation of TFSI into graphite, graphene layer donates electrons due to a decrease of its average L\"{o}wdin charge per atom from 3.962 in graphite to 3.936 (\ce{C50}) $\sim$ 3.902 (\ce{C18}) in GICs, while TFSI molecule plays a role of electron acceptor due to an increase of its L\"{o}wdin charge from 5.973 in the isolated state to 6.036 in GICs. As increasing the cell size of graphene sheet in the cases of GICs, the amount of donated electron by graphene layer decreases due to an increase of L\"{o}wdin charge per carbon atom, but the average charge of TFSI molecule per atom keeps a constant, which are in good agreement with the intuitive observation of electronic density differences. Inside the TFSI molecule, only the charge of S atom decreases upon the intercalation, while those of other atoms increase, indicating that S atoms also donate electrons, as already mentioned in the analysis of molecular orbitals.

\section{Conclusions}
With first-principles calculations using the PBE + vdW-DF2 functional, we studied the intercalation of TFSI anion into graphite, aiming to identify the atomistic structures, energetics, electrochemical properties, and electronic charge transfer of TFSI-C$_n$ compounds for dual ion battery application. We made modelling of these compounds using various sizes of graphene unit cells from $(3\times3)$ to $(6\times6)$, producing the number of carbon atoms $n$ from 18 to 72, and checked their formation energies, clarifying the linear increase of formation energy as increasing the cell size and thermodynamic instability of GICs over $n$ = 60 due to positive formation energies. Together with the decrease of interlayer binding energy as well, this indicates that at low concentration of TFSI intercalate TFSI-C$_n$ compounds can not be formed. As increasing the carbon atom number in TFSI-C$_n$ compounds, the interlayer distance was found to decrease from 7.89 \AA~(\ce{C18}) to 7.79 \AA~(\ce{C50}), and accordingly the volume expansion ratio relative to graphite also decreases from 130 \% to 127\% in slight underestimation compared with experiment. The average electrode potential during the charge process was determined to be ranged from 3.8 V to 3.0 V at the specific capacity range from 30 mAh g$^{-1}$ to 54 mAh g$^{-1}$ in reasonable agreement with experiment. In particular, the activation barrier for TFSI migration inside graphite was calculated to be quite low like under 50 meV, suggesting the very fast charging time. The analysis of DOS indicates that these compounds can be electron conductors and the conductivity seems to decrease as increasing the concentration of intercalate. Through the analysis of electronic density difference and atomic charges, the graphene layer and TFSI molecule in these compounds play role of electron donor and acceptor, respectively. The calculation results clarify the mechanism of TFSI-C$_n$ formation and reveal new prospects for developing graphite based cathode materials of alkali ion batteries.

\section*{Acknowledgments}
This work was supported partially from the State Committee of Science and Technology, Democratic People's Republic of Korea, under the fundamental research project ``Design of Innovative Functional Materials for Energy and Environmental Application'' (No. 2016-20). The calculations in this work were carried out on the HP Blade System C7000 (HP BL460c) that is owned by Faculty of Materials Science, Kim Il Sung University.

\section*{Appendix A. Supplementary data}
Supplementary data related to this article can be found at URL.

\section*{Notes}
The authors declare no competing financial interest.

\bibliographystyle{elsarticle-num-names}
\bibliography{Reference}

\begin{thebibliography}{50}
\providecommand{\natexlab}[1]{#1}
\providecommand{\url}[1]{\texttt{#1}}
\providecommand{\urlprefix}{URL }
\expandafter\ifx\csname urlstyle\endcsname\relax
  \providecommand{\doi}[1]{doi:\discretionary{}{}{}#1}\else
  \providecommand{\doi}[1]{doi:\discretionary{}{}{}\begingroup
  \urlstyle{rm}\url{#1}\endgroup}\fi
\providecommand{\bibinfo}[2]{#2}

\bibitem[{Par(2015)}]{Paris}
\bibinfo{title}{EUROPEAN COMMISSION Climate Action. Annual Conference of
  Parties (COP) at COP21 at Stade de France (Gate E) in Paris},
  \bibinfo{note}{http://ec.europa.eu/clima/policies/international/negotiations%
/paris/index\_en.htm}, \bibinfo{year}{2015}.

\bibitem[{Tarascon(2010)}]{Tarascon10}
\bibinfo{author}{J.-M. Tarascon}, \bibinfo{title}{Is lithium the new gold?},
  \bibinfo{journal}{Nature Chemistry} \bibinfo{volume}{2}
  (\bibinfo{year}{2010}) \bibinfo{pages}{510}.

\bibitem[{Goodenough and Kim(2010)}]{Goodenough10}
\bibinfo{author}{J.~B. Goodenough}, \bibinfo{author}{Y.~Kim},
  \bibinfo{title}{Challenges for Rechargeable Li Batteries},
  \bibinfo{journal}{Chem. Mater.} \bibinfo{volume}{22} (\bibinfo{year}{2010})
  \bibinfo{pages}{587--603}.

\bibitem[{Zhang et~al.(2017)Zhang, Mao, Li, and Chen}]{HZhang}
\bibinfo{author}{H.~Zhang}, \bibinfo{author}{C.~Mao}, \bibinfo{author}{J.~Li},
  \bibinfo{author}{R.~Chen}, \bibinfo{title}{Advances in electrode materials
  for Li-based rechargeable batteries}, \bibinfo{journal}{RSC Adv.}
  \bibinfo{volume}{7} (\bibinfo{year}{2017}) \bibinfo{pages}{33789--33811}.

\bibitem[{Hwang et~al.(2017)Hwang, Myung, and Sun}]{JYHwang}
\bibinfo{author}{J.-Y. Hwang}, \bibinfo{author}{S.-T. Myung},
  \bibinfo{author}{Y.-K. Sun}, \bibinfo{title}{Sodium-ion batteries: present
  and future}, \bibinfo{journal}{Chem. Soc. Rev.} \bibinfo{volume}{46}
  (\bibinfo{year}{2017}) \bibinfo{pages}{3529--3614}.

\bibitem[{Yabuuchi et~al.(2014)Yabuuchi, Kubota, Dahbi, and
  Komaba}]{Yabuuchi14}
\bibinfo{author}{N.~Yabuuchi}, \bibinfo{author}{K.~Kubota},
  \bibinfo{author}{M.~Dahbi}, \bibinfo{author}{S.~Komaba},
  \bibinfo{title}{Research Development on Sodium-Ion Batteries},
  \bibinfo{journal}{Chem. Rev.} \bibinfo{volume}{114} (\bibinfo{year}{2014})
  \bibinfo{pages}{11636--11682}.

\bibitem[{Sawicki and Shaw(2015)}]{Sawicki}
\bibinfo{author}{M.~Sawicki}, \bibinfo{author}{L.~L. Shaw},
  \bibinfo{title}{Advances and challenges of sodium ion batteries as post
  lithium ion batteries}, \bibinfo{journal}{RSC Adv.} \bibinfo{volume}{5}
  (\bibinfo{year}{2015}) \bibinfo{pages}{53129--53154}.

\bibitem[{Pramudita et~al.(2017)Pramudita, Sehrawat, Goonetilleke, and
  Sharma}]{Pramudita}
\bibinfo{author}{J.~C. Pramudita}, \bibinfo{author}{D.~Sehrawat},
  \bibinfo{author}{D.~Goonetilleke}, \bibinfo{author}{N.~Sharma},
  \bibinfo{title}{An Initial Review of the Status of Electrode Materials for
  Potassium-Ion Batteries}, \bibinfo{journal}{Adv. Energy Mater.}
  (\bibinfo{year}{2017}) \bibinfo{pages}{1602911}.

\bibitem[{Ji et~al.(2017)Ji, Zhang, Song, and Tang}]{BJi}
\bibinfo{author}{B.~Ji}, \bibinfo{author}{F.~Zhang}, \bibinfo{author}{X.~Song},
  \bibinfo{author}{Y.~Tang}, \bibinfo{title}{A Novel Potassium-Ion-Based
  Dual-Ion Battery}, \bibinfo{journal}{Adv. Mater.} \bibinfo{volume}{29}
  (\bibinfo{year}{2017}) \bibinfo{pages}{1700519--1700525}.

\bibitem[{Dresselhaus and Dresselhaus(2002)}]{Dresselhaus}
\bibinfo{author}{M.~S. Dresselhaus}, \bibinfo{author}{G.~Dresselhaus},
  \bibinfo{title}{Intercalation compounds of graphite}, \bibinfo{journal}{Adv.
  Phys.} \bibinfo{volume}{51} (\bibinfo{year}{2002}) \bibinfo{pages}{1--186}.

\bibitem[{Moriwake et~al.(2017)Moriwake, Kuwabara, Fisher, and
  Ikuhara}]{Moriwake}
\bibinfo{author}{H.~Moriwake}, \bibinfo{author}{A.~Kuwabara},
  \bibinfo{author}{C.~A.~J. Fisher}, \bibinfo{author}{Y.~Ikuhara},
  \bibinfo{title}{Why is sodium-intercalated graphite unstable?},
  \bibinfo{journal}{RSC Adv.} \bibinfo{volume}{7} (\bibinfo{year}{2017})
  \bibinfo{pages}{36550--587}.

\bibitem[{Nobuhara et~al.(2013)Nobuhara, Nakayama, Nose, Nakanishi, and
  Iba}]{Nobuhara}
\bibinfo{author}{K.~Nobuhara}, \bibinfo{author}{H.~Nakayama},
  \bibinfo{author}{M.~Nose}, \bibinfo{author}{S.~Nakanishi},
  \bibinfo{author}{H.~Iba}, \bibinfo{title}{First-principles study of alkali
  metal-graphite intercalation compounds}, \bibinfo{journal}{J. Power Sources}
  \bibinfo{volume}{243} (\bibinfo{year}{2013}) \bibinfo{pages}{585--587}.

\bibitem[{Wang et~al.(2014)Wang, Selbach, and Grande}]{ZWang}
\bibinfo{author}{Z.~Wang}, \bibinfo{author}{S.~M. Selbach},
  \bibinfo{author}{T.~Grande}, \bibinfo{title}{Van der Waals density functional
  study of the energetics of alkali metal intercalation in graphite},
  \bibinfo{journal}{RSC Adv.} \bibinfo{volume}{4} (\bibinfo{year}{2014})
  \bibinfo{pages}{4069--4079}.

\bibitem[{Ri et~al.(2016)Ri, Yu, Kim, Hong, Jong, and Ri}]{yucj06}
\bibinfo{author}{G.-C. Ri}, \bibinfo{author}{C.-J. Yu}, \bibinfo{author}{J.-S.
  Kim}, \bibinfo{author}{S.-N. Hong}, \bibinfo{author}{U.-G. Jong},
  \bibinfo{author}{M.-H. Ri}, \bibinfo{title}{First-principles study of ternary
  graphite compounds cointercalated with alkali atoms (Li, Na, and K) and
  alkylamines towards alkali ion battery applications}, \bibinfo{journal}{J.
  Power Sources} \bibinfo{volume}{324} (\bibinfo{year}{2016})
  \bibinfo{pages}{758--765}.

\bibitem[{Vaalma et~al.(2016)Vaalma, Giffin, Buchholz, and Passerini}]{Vaalma}
\bibinfo{author}{C.~Vaalma}, \bibinfo{author}{G.~A. Giffin},
  \bibinfo{author}{D.~Buchholz}, \bibinfo{author}{S.~Passerini},
  \bibinfo{title}{Non-Aqueous K-Ion Battery Based on Layered \ce{K_{0.3}MnO2}
  and Hard Carbon/Carbon Black}, \bibinfo{journal}{J. Electrochem. Soc.}
  \bibinfo{volume}{163} (\bibinfo{year}{2016}) \bibinfo{pages}{A1295--A1299}.

\bibitem[{Ren et~al.(2017)Ren, Zhao, McCulloch, and Wu}]{XRen}
\bibinfo{author}{X.~Ren}, \bibinfo{author}{Q.~Zhao}, \bibinfo{author}{W.~D.
  McCulloch}, \bibinfo{author}{Y.~Wu}, \bibinfo{title}{\ce{MoS2} as a long-life
  host material for potassium ion intercalation}, \bibinfo{journal}{Nano Res.}
  \bibinfo{volume}{10} (\bibinfo{year}{2017}) \bibinfo{pages}{1313--1321}.

\bibitem[{Han et~al.(2017)Han, Li, Liu, Wang, Zhang, Hu, Dai, and Xu}]{JHan}
\bibinfo{author}{J.~Han}, \bibinfo{author}{G.-N. Li}, \bibinfo{author}{F.~Liu},
  \bibinfo{author}{M.~Wang}, \bibinfo{author}{Y.~Zhang},
  \bibinfo{author}{L.~Hu}, \bibinfo{author}{C.~Dai}, \bibinfo{author}{M.~Xu},
  \bibinfo{title}{Investigation of \ce{K3V2(PO4)3}/C nanocomposites as
  high-potential cathode materials for potassium-ion batteries},
  \bibinfo{journal}{Chem. Commun.} \bibinfo{volume}{53} (\bibinfo{year}{2017})
  \bibinfo{pages}{1805--1808}.

\bibitem[{Recham et~al.(2012)Recham, l.~Rousse, Sougrati, Chotard, Frayret,
  Mariyappan, Melot, Jumas, and Tarascon}]{Recham}
\bibinfo{author}{N.~Recham}, \bibinfo{author}{G.~l.~Rousse},
  \bibinfo{author}{M.~T. Sougrati}, \bibinfo{author}{J.-N. Chotard},
  \bibinfo{author}{C.~Frayret}, \bibinfo{author}{S.~Mariyappan},
  \bibinfo{author}{B.~C. Melot}, \bibinfo{author}{J.-C. Jumas},
  \bibinfo{author}{J.-M. Tarascon}, \bibinfo{title}{Preparation and
  Characterization of a Stable \ce{FeSO4F}-Based Framework for Alkali Ion
  Insertion Electrodes}, \bibinfo{journal}{Chem. Mater.} \bibinfo{volume}{24}
  (\bibinfo{year}{2012}) \bibinfo{pages}{4363--4370}.

\bibitem[{Wang et~al.(2007)Wang, Zaghib, Guerfi, Bazito, Torresi, and
  Dahn}]{YWang}
\bibinfo{author}{Y.~Wang}, \bibinfo{author}{K.~Zaghib},
  \bibinfo{author}{A.~Guerfi}, \bibinfo{author}{F.~F.~C. Bazito},
  \bibinfo{author}{R.~M. Torresi}, \bibinfo{author}{J.~R. Dahn},
  \bibinfo{title}{Accelerating rate calorimetry studies of the reactions
  between ionic liquids and charged lithium ion battery electrode materials},
  \bibinfo{journal}{Electrochim. Acta} \bibinfo{volume}{52}
  (\bibinfo{year}{2007}) \bibinfo{pages}{6346--6352}.

\bibitem[{Abu-Lebdeh et~al.(2006)Abu-Lebdeh, Abouimrane, Alarco, and
  Armand}]{Lebdeh}
\bibinfo{author}{Y.~Abu-Lebdeh}, \bibinfo{author}{A.~Abouimrane},
  \bibinfo{author}{P.-J. Alarco}, \bibinfo{author}{M.~Armand},
  \bibinfo{title}{Ionic liquid and plastic crystalline phases of pyrazolium
  imide salts as electrolytes for rechargeable lithium-ion batteries},
  \bibinfo{journal}{J. Power Sources} \bibinfo{volume}{154}
  (\bibinfo{year}{2006}) \bibinfo{pages}{255--261}.

\bibitem[{Carlin et~al.(1994)Carlin, Long, Fuller, and Trulove}]{Carlin}
\bibinfo{author}{R.~T. Carlin}, \bibinfo{author}{H.~C.~D. Long},
  \bibinfo{author}{J.~Fuller}, \bibinfo{author}{P.~C. Trulove},
  \bibinfo{title}{{Dual Intercalating Molten Electrolyte Batteries}},
  \bibinfo{journal}{J. Electrochem. Soc.} \bibinfo{volume}{141}
  (\bibinfo{year}{1994}) \bibinfo{pages}{L73--L76}.

\bibitem[{Seel and Dahn(2000)}]{Seel}
\bibinfo{author}{J.~A. Seel}, \bibinfo{author}{J.~R. Dahn},
  \bibinfo{title}{{Electrochemical Intercalation of PF$_6$ into Graphite}},
  \bibinfo{journal}{J. Electrochem. Soc.} \bibinfo{volume}{147}
  (\bibinfo{year}{2000}) \bibinfo{pages}{892--898}.

\bibitem[{Schmuelling et~al.(2013)Schmuelling, Placke, Kloepsch, Fromm, Meyer,
  Passerini, and Winter}]{Schmuelling}
\bibinfo{author}{G.~Schmuelling}, \bibinfo{author}{T.~Placke},
  \bibinfo{author}{R.~Kloepsch}, \bibinfo{author}{O.~Fromm},
  \bibinfo{author}{H.-W. Meyer}, \bibinfo{author}{S.~Passerini},
  \bibinfo{author}{M.~Winter}, \bibinfo{title}{{X-Ray Diffraction Studies of
  the Electrochemical Intercalation of Bis(trifluoromethanesulfonyl)imide
  Anions into Graphite for Dual-Ion Cells}}, \bibinfo{journal}{J. Power
  Sources} \bibinfo{volume}{239} (\bibinfo{year}{2013})
  \bibinfo{pages}{563--571}.

\bibitem[{Beltrop et~al.(2016)Beltrop, Meister, Klein, Heckmann, Gr\"{u}nebaum,
  Wiemh\"{o}fer, Winter, and Placke}]{Beltrop2}
\bibinfo{author}{K.~Beltrop}, \bibinfo{author}{P.~Meister},
  \bibinfo{author}{S.~Klein}, \bibinfo{author}{A.~Heckmann},
  \bibinfo{author}{M.~Gr\"{u}nebaum}, \bibinfo{author}{H.-D. Wiemh\"{o}fer},
  \bibinfo{author}{M.~Winter}, \bibinfo{author}{T.~Placke}, \bibinfo{title}{New
  Insights into the Intercalation Behavior of Anions into a Graphite-Based
  Positive Electrode for Dual-Ion Batteries}, \bibinfo{journal}{Electrochim.
  Acta} \bibinfo{volume}{209} (\bibinfo{year}{2016}) \bibinfo{pages}{44--55}.

\bibitem[{Rothermel et~al.(2014)Rothermel, Meister, Schmuelling, Fromm, Meyer,
  Nowak, Winter, and Placke}]{Rothermel}
\bibinfo{author}{S.~Rothermel}, \bibinfo{author}{P.~Meister},
  \bibinfo{author}{G.~Schmuelling}, \bibinfo{author}{O.~Fromm},
  \bibinfo{author}{H.~W. Meyer}, \bibinfo{author}{S.~Nowak},
  \bibinfo{author}{M.~Winter}, \bibinfo{author}{T.~Placke},
  \bibinfo{title}{Dual-Graphite Cells based on the Reversible Intercalation of
  Bis (trifluoromethanesulfonyl) imide Anions from an Ionic Liquid
  Electrolyte}, \bibinfo{journal}{Energy Environ. Sci.} \bibinfo{volume}{7}
  (\bibinfo{year}{2014}) \bibinfo{pages}{3412--3423}.

\bibitem[{Meister et~al.(2014)Meister, Siozios, Reiter, Klamor, Rothermel,
  Fromm, Meyer, Winter, and Placke}]{Meister}
\bibinfo{author}{P.~Meister}, \bibinfo{author}{V.~Siozios},
  \bibinfo{author}{J.~Reiter}, \bibinfo{author}{S.~Klamor},
  \bibinfo{author}{S.~Rothermel}, \bibinfo{author}{O.~Fromm},
  \bibinfo{author}{H.~W. Meyer}, \bibinfo{author}{M.~Winter},
  \bibinfo{author}{T.~Placke}, \bibinfo{title}{Dual-Ion Cells based on the
  Electrochemical Intercalations of Asymmetric Fluorosulfonyl
  (trifluoromethanesulfonyl) imide Anions into Graphite},
  \bibinfo{journal}{Electrochim. Acta} \bibinfo{volume}{130}
  (\bibinfo{year}{2014}) \bibinfo{pages}{625--633}.

\bibitem[{Santhanam and Noel(1998)}]{Santhanam}
\bibinfo{author}{R.~Santhanam}, \bibinfo{author}{M.~Noel},
  \bibinfo{title}{{Electrochemical Intercalation of Cationic and Anionic
  Species from a Lithium Perchlorate-Propylene Carbonate System--a
  Rocking-Chair Type of Dual-Intercalation System}}, \bibinfo{journal}{J. Power
  Sources} \bibinfo{volume}{76} (\bibinfo{year}{1998})
  \bibinfo{pages}{147--152}.

\bibitem[{Fujii et~al.(2006)Fujii, Fujimori, Takamuku, Kanzaki, Umebayashi, and
  Ishiguro}]{Fujii}
\bibinfo{author}{K.~Fujii}, \bibinfo{author}{T.~Fujimori},
  \bibinfo{author}{T.~Takamuku}, \bibinfo{author}{R.~Kanzaki},
  \bibinfo{author}{Y.~Umebayashi}, \bibinfo{author}{S.~Ishiguro},
  \bibinfo{title}{{Conformational Equilibrium of Bis(trifluoromethanesulfonyl)
  Imide Anion of a Room-Temperature Ionic Liquid: Raman Spectroscopic Study and
  DFT Calculations}}, \bibinfo{journal}{J. Phys. Chem. B} \bibinfo{volume}{110}
  (\bibinfo{year}{2006}) \bibinfo{pages}{8179--8183}.

\bibitem[{Umebayashi et~al.(2007)Umebayashi, Mitsugi, Fukuda, Fujimori, Fujii,
  Kanzaki, Takeuchi, and Ishiguro}]{Umebayashi}
\bibinfo{author}{Y.~Umebayashi}, \bibinfo{author}{T.~Mitsugi},
  \bibinfo{author}{S.~Fukuda}, \bibinfo{author}{T.~Fujimori},
  \bibinfo{author}{K.~Fujii}, \bibinfo{author}{R.~Kanzaki},
  \bibinfo{author}{M.~Takeuchi}, \bibinfo{author}{S.~Ishiguro},
  \bibinfo{title}{{Lithium Ion Solvation in Room-Temperature Ionic Liquids
  Involving Bis(trifluoromethanesulfonyl) Imide Anion Studied by Raman
  Spectroscopy and DFT Calculations}}, \bibinfo{journal}{J. Phys. Chem. B}
  \bibinfo{volume}{111} (\bibinfo{year}{2007}) \bibinfo{pages}{13028--13032}.

\bibitem[{Henderson and Passerini(2004)}]{Henderson}
\bibinfo{author}{W.~A. Henderson}, \bibinfo{author}{S.~Passerini},
  \bibinfo{title}{Phase Behavior of Ionic Liquid-LiX Mixtures: Pyrrolidinium
  Cations and TFSI$^-$ Anions}, \bibinfo{journal}{Chem. Mater.}
  \bibinfo{volume}{16} (\bibinfo{year}{2004}) \bibinfo{pages}{2881--2885}.

\bibitem[{Henderson et~al.(2006)Henderson, Herstedt, Young, Passerini, Long,
  and Trulove}]{Henderson2}
\bibinfo{author}{W.~A. Henderson}, \bibinfo{author}{M.~Herstedt},
  \bibinfo{author}{V.~G. Young}, \bibinfo{author}{S.~Passerini},
  \bibinfo{author}{H.~C.~D. Long}, \bibinfo{author}{P.~C. Trulove},
  \bibinfo{title}{New Disordering Mode for TFSI$^-$ Anions: The Nonequilibrium,
  Plastic Crystalline Structure of Et$_4$NTFSI}, \bibinfo{journal}{Inorg.
  Chem.} \bibinfo{volume}{45} (\bibinfo{year}{2006})
  \bibinfo{pages}{1412--1414}.

\bibitem[{Bhatt et~al.(2006)Bhatt, Duffy, Collison, May, and Lewin}]{Bhatt}
\bibinfo{author}{A.~I. Bhatt}, \bibinfo{author}{N.~W. Duffy},
  \bibinfo{author}{D.~Collison}, \bibinfo{author}{I.~May},
  \bibinfo{author}{R.~G. Lewin}, \bibinfo{title}{Cyclic Voltammetry of Th(IV)
  in the Room-Temperature Ionic Liquid [Me$_3$N$^n$Bu][\ce{N(SO2CF3)2}]},
  \bibinfo{journal}{Inorg. Chem.} \bibinfo{volume}{45} (\bibinfo{year}{2006})
  \bibinfo{pages}{1677--1682}.

\bibitem[{Siqueira and Ribeiro(2007)}]{Siqueira}
\bibinfo{author}{L.~J.~A. Siqueira}, \bibinfo{author}{M.~C.~C. Ribeiro},
  \bibinfo{title}{{Molecular Dynamics Simulation of the Ionic Liquid
  N-Ethyl-N,N-dimethyl-N-(2-methoxyethyl)ammonium
  Bis(trifluoromethanesulfonyl)imide}}, \bibinfo{journal}{J. Phys. Chem. B}
  \bibinfo{volume}{111} (\bibinfo{year}{2007}) \bibinfo{pages}{11776--11785}.

\bibitem[{Borodin and Smith(2006)}]{Borodin1}
\bibinfo{author}{O.~Borodin}, \bibinfo{author}{G.~D. Smith},
  \bibinfo{title}{{Structure and Dynamics of N-Methyl-N-propylpyrrolidinium
  Bis(trifluoromethanesulfonyl)imide Ionic Liquid from Molecular Dynamics
  Simulations}}, \bibinfo{journal}{J. Phys. Chem. B} \bibinfo{volume}{110}
  (\bibinfo{year}{2006}) \bibinfo{pages}{11481--11490}.

\bibitem[{Borodin et~al.(2006)Borodin, Smith, and Henderson}]{Borodin2}
\bibinfo{author}{O.~Borodin}, \bibinfo{author}{G.~D. Smith},
  \bibinfo{author}{W.~Henderson}, \bibinfo{title}{{Li$^+$ Cation Environment,
  Transport, and Mechanical Properties of the LiTFSI Doped
  N-Methyl-N-alkylpyrrolidinium$^+$TFSI$^-$ Ionic Liquids}},
  \bibinfo{journal}{J. Phys. Chem. B} \bibinfo{volume}{110}
  (\bibinfo{year}{2006}) \bibinfo{pages}{16879--16886}.

\bibitem[{Nicotera et~al.(2005)Nicotera, Oliviero, Henderson, Appetecchi, and
  Passerini}]{Nicotera}
\bibinfo{author}{I.~Nicotera}, \bibinfo{author}{C.~Oliviero},
  \bibinfo{author}{W.~A. Henderson}, \bibinfo{author}{G.~B. Appetecchi},
  \bibinfo{author}{S.~Passerini}, \bibinfo{title}{{NMR Investigation of Ionic
  Liquid-LiX Mixtures: Pyrrolidinium Cations and TFSI$^-$ Anions}},
  \bibinfo{journal}{J. Phys. Chem. B} \bibinfo{volume}{109}
  (\bibinfo{year}{2005}) \bibinfo{pages}{22814--22819}.

\bibitem[{Beltrop et~al.(2017)Beltrop, Beuker, Heckmann, Winter, and
  Placke}]{Beltrop1}
\bibinfo{author}{K.~Beltrop}, \bibinfo{author}{S.~Beuker},
  \bibinfo{author}{A.~Heckmann}, \bibinfo{author}{M.~Winter},
  \bibinfo{author}{T.~Placke}, \bibinfo{title}{Alternative electrochemical
  energy storage: potassium-based dual-graphite batteries},
  \bibinfo{journal}{Energy Environ. Sci.} \bibinfo{volume}{10}
  (\bibinfo{year}{2017}) \bibinfo{pages}{DOI: 10.1039/c7ee01535f}.

\bibitem[{Tasaki(2014)}]{Tasaki}
\bibinfo{author}{K.~Tasaki}, \bibinfo{title}{{Density Functional Theory Study
  on Structural and Energetic Characteristics of Graphite Intercalation
  Compounds}}, \bibinfo{journal}{J. Phys. Chem. C} \bibinfo{volume}{118}
  (\bibinfo{year}{2014}) \bibinfo{pages}{1443--1450}.

\bibitem[{{P. Giannozzi and S. Baroni and N. Bonini and M. Calandra and R. Car,
  {\it et al.}}(2009)}]{QE}
\bibinfo{author}{{P. Giannozzi and S. Baroni and N. Bonini and M. Calandra and
  R. Car, {\it et al.}}}, \bibinfo{title}{{QUANTUM ESPRESSO}: a modular and
  open-source software project for quantum simulations of materials},
  \bibinfo{journal}{J. Phys.:Condens. Matter} \bibinfo{volume}{21}
  (\bibinfo{year}{2009}) \bibinfo{pages}{395502}.

\bibitem[{Perdew et~al.(1996)Perdew, Burke, and Ernzerhof}]{PBE}
\bibinfo{author}{J.~P. Perdew}, \bibinfo{author}{K.~Burke},
  \bibinfo{author}{M.~Ernzerhof}, \bibinfo{title}{{Generalized Gradient
  Approximation Made Simple}}, \bibinfo{journal}{Phys. Rev. Lett.}
  \bibinfo{volume}{77} (\bibinfo{year}{1996}) \bibinfo{pages}{3865}.

\bibitem[{Lee et~al.(2010)Lee, Murray, Kong, Lundqvist, and Langreth}]{vdwDF2}
\bibinfo{author}{K.~Lee}, \bibinfo{author}{E.~D. Murray},
  \bibinfo{author}{L.~Kong}, \bibinfo{author}{B.~I. Lundqvist},
  \bibinfo{author}{D.~C. Langreth}, \bibinfo{title}{High-accuracy van der Waals
  density functional}, \bibinfo{journal}{Phys. Rev. B} \bibinfo{volume}{82}
  (\bibinfo{year}{2010}) \bibinfo{pages}{081101(R)}.

\bibitem[{Yu et~al.(2017)Yu, Ri, Choe, Ri, Kye, and Kim}]{yucj14}
\bibinfo{author}{C.-J. Yu}, \bibinfo{author}{S.-B. Ri}, \bibinfo{author}{S.-H.
  Choe}, \bibinfo{author}{G.-C. Ri}, \bibinfo{author}{Y.-H. Kye},
  \bibinfo{author}{S.-C. Kim}, \bibinfo{title}{Ab initio study of sodium
  cointercalation with diglyme molecule into graphite},
  \bibinfo{journal}{Electrochim. Acta} \bibinfo{volume}{253}
  (\bibinfo{year}{2017}) \bibinfo{pages}{589--598}.

\bibitem[{Henkelman et~al.(2000)Henkelman, Uberuaga, and J\'{o}nsson}]{NEB}
\bibinfo{author}{G.~Henkelman}, \bibinfo{author}{B.~P. Uberuaga},
  \bibinfo{author}{H.~J\'{o}nsson}, \bibinfo{title}{A climbing image nudged
  elastic band method for finding saddle points and minimum energy paths},
  \bibinfo{journal}{J. Chem. Phys.} \bibinfo{volume}{113}
  (\bibinfo{year}{2000}) \bibinfo{pages}{9901--9904}.

\bibitem[{Yu et~al.(2014)Yu, Ri, Jong, Choe, and Cha}]{yucj04}
\bibinfo{author}{C.-J. Yu}, \bibinfo{author}{G.-C. Ri}, \bibinfo{author}{U.-G.
  Jong}, \bibinfo{author}{Y.-G. Choe}, \bibinfo{author}{S.-J. Cha},
  \bibinfo{title}{Refined phase coexistence line between graphite and diamond
  from density-functional theory and van der {Waals} correction},
  \bibinfo{journal}{Physica B} \bibinfo{volume}{434} (\bibinfo{year}{2014})
  \bibinfo{pages}{185--193}.

\bibitem[{Tsai et~al.(2015)Tsai, Chung, Linb, and Yamada}]{Tsai}
\bibinfo{author}{P.-C. Tsai}, \bibinfo{author}{S.-C. Chung},
  \bibinfo{author}{S.-K. Linb}, \bibinfo{author}{A.~Yamada}, \bibinfo{title}{Ab
  initio study of sodium intercalation into disordered carbon},
  \bibinfo{journal}{J. Mater. Chem. A} \bibinfo{volume}{3}
  (\bibinfo{year}{2015}) \bibinfo{pages}{9763--9768}.

\bibitem[{Boettiger(1997)}]{boettiger}
\bibinfo{author}{J.~C. Boettiger}, \bibinfo{title}{All-electron full-potential
  calculation of the electronic band structure, elastic constants, and equation
  of state for graphite}, \bibinfo{journal}{Phys. Rev. B} \bibinfo{volume}{55}
  (\bibinfo{year}{1997}) \bibinfo{pages}{11202}.

\bibitem[{Zacharia et~al.(2004)Zacharia, Ulbricht, and Hertel}]{graphitexf1}
\bibinfo{author}{R.~Zacharia}, \bibinfo{author}{H.~Ulbricht},
  \bibinfo{author}{T.~Hertel}, \bibinfo{title}{Interlayer Cohesive Energy of
  Graphite from Thermal Desorption of Polyaromatic Hydrocarbons},
  \bibinfo{journal}{Phys. Rev. B} \bibinfo{volume}{69} (\bibinfo{year}{2004})
  \bibinfo{pages}{155406--155412}.

\bibitem[{Benedict et~al.(1998)Benedict, Chopra, Cohen, Zettl, Louie, and
  Crespi}]{graphitexf2}
\bibinfo{author}{L.~X. Benedict}, \bibinfo{author}{N.~G. Chopra},
  \bibinfo{author}{M.~L. Cohen}, \bibinfo{author}{A.~Zettl},
  \bibinfo{author}{S.~G. Louie}, \bibinfo{author}{V.~H. Crespi},
  \bibinfo{title}{Microscopic Determination of the Interlayer Binding Energy in
  Graphite}, \bibinfo{journal}{Chem. Phys. Lett.} \bibinfo{volume}{286}
  (\bibinfo{year}{1998}) \bibinfo{pages}{490--496}.

\bibitem[{Dappe et~al.(2006)Dappe, Basanta, Flores, and Ortega}]{graphitexf3}
\bibinfo{author}{Y.~J. Dappe}, \bibinfo{author}{M.~A. Basanta},
  \bibinfo{author}{F.~Flores}, \bibinfo{author}{J.~Ortega},
  \bibinfo{title}{Reflection Second Harmonic Generation on a z-Cut Congruent
  Lithium Niobate Crystal}, \bibinfo{journal}{Phys. Rev. B}
  \bibinfo{volume}{74} (\bibinfo{year}{2006}) \bibinfo{pages}{205424--205430}.

\bibitem[{Otani and Sugino(2006)}]{esm}
\bibinfo{author}{M.~Otani}, \bibinfo{author}{O.~Sugino},
  \bibinfo{title}{First-principles calculations of charged surfaces and
  interfaces: A plane-wave nonrepeated slab approach}, \bibinfo{journal}{Phys.
  Rev. B} \bibinfo{volume}{73} (\bibinfo{year}{2006}) \bibinfo{pages}{115407}.

\end{thebibliography}

\end{document}